\documentclass{amsart}

\usepackage{graphics}

\usepackage[ps,matrix,arrow,poly,graph,curve]{xy}
\UsePSspecials{dvips}                      

\labelmargin-{1.2pt}

\usepackage{amsmath}

\DeclareMathOperator{\Spin}{Spin}
\DeclareMathOperator{\SU}{SU}
\DeclareMathOperator{\TetW}{Tet}
\usepackage{xspace}
\newcommand{\Spinf}{\ensuremath{\Spin(4)}\xspace}
\newcommand{\SUt}{\ensuremath{\SU(2)}\xspace}

\newcommand{\cxymatrix}[1]{\vcenter{\xymatrix{#1}}}

\newcommand{\sixJ}[1]{\ensuremath{\left\{\begin{matrix}#1\end{matrix}\right\}}}
\newcommand{\Tet}[1]{\ensuremath{\TetW\left[\begin{matrix}#1\end{matrix}\right]}}

\newcommand{\DoubleYOld}[5]
{  
\cxymatrix{\ar@{-}[dr]^{#1} & & \ar@{-}[dl]_{#2} \\  
 & *{\bullet} \ar@{-}[d]^{#5} \\
 & *{\bullet} \\
\ar@{-}[ur]_{#3} & & \ar@{-}[ul]^{#4}}  
}  

\newcommand{\DoubleY}[5] 
{
\begin{xy} 
\xygraph{!{<1.6pc,0pc>:<0pc,1.8pc>::0}
   [u(0.7)] *{\bullet}
   -@-_>>{#1} [ul] [dr]
   -@-^>>{#2} [ur] [dl] 
   -@-^{#5} [d(1.4)] *{\bullet}
   -@-^>>{#3} [dl] [ur] 
   -@-_>>{#4} [dr] [ul]}
\end{xy}
}

\newcommand{\DoubleYhorOld}[5] 
{  
\cxymatrix{\ar@{-}[dr]^{#1} & & & \ar@{-}[dl]_{#2} \\  
 & *{\bullet} \ar@{-}[r]^{#5}& *{\bullet} &\\  
\ar@{-}[ur]_{#3} & & & \ar@{-}[ul]^{#4}}  
}  

\newcommand{\DoubleYhor}[5] 
{
\begin{xy} 
\xygraph{!{<0pc,1.8pc>:<1.6pc,0pc>::0}
   [u(0.7)] *{\bullet}
   -@-_>>{#4} [ul] [dr]
   -@-^>>{#2} [ur] [dl] 
   -@-_{#5} [d(1.4)] *{\bullet}
   -@-^>>{#3} [dl] [ur] 
   -@-_>>{#1} [dr] [ul]}
\end{xy}
}

\newcommand{\FourX}[4]
{
\begin{xy} 
\xygraph{!{<1.6pc,0pc>:<0pc,2.2pc>::0}
  *{\bullet}
   -@-_>>{#1} [ul] [dr]
   -@-^>>{#2} [ur] [dl] 
   -@-^>>{#3} [dl] [ur] 
   -@-_>>{#4} [dr] [ul]}
\end{xy}
}

\newcommand{\xyphi}[4]
{
\begin{xy}
\xygraph{!{<1pc,0pc>:}
   [u] *{\bullet}
   -@-_>>{#1} [u] [d]
   -@/_.6pc/_{#2} [dd] [uu]
   -@/^.6pc/^{#3} [dd] *{\bullet}
   -@-^>>{#4} [d] [u]}
\end{xy}
}

\newcommand{\order}{{\mathcal O}}

\newcommand{\Center}[1]{\begin{matrix}#1\end{matrix}}

\begin{document}

\title{An efficient algorithm for the Riemannian $10j$ symbols}
\author{J. Daniel Christensen}
\address{Dept of Mathematics\\
University of Western Ontario\\
London, Ontario, Canada}
\email{jdc@uwo.ca}
\author{Greg Egan}
\email{gregegan@netspace.net.au}
\thanks{The authors would like to thank John Baez for 
many useful conversations about the material in this paper.}
\date{January 23, 2002}

\begin{abstract}
\noindent
The $10j$ symbol is a spin network that appears in the partition function for
the Barrett-Crane model of Riemannian quantum gravity.  Elementary methods of
calculating the $10j$ symbol require $\order(j^9)$ or more operations and 
$\order(j^2)$ or more space, where $j$ is the average spin.
We present an algorithm that computes the $10j$ symbol using $\order(j^5)$ operations
and $\order(j^{2})$ space, and a variant that uses $\order(j^6)$ operations and
a constant amount of space.
An implementation has been made available on the web.
\end{abstract}

\maketitle

\section{Introduction}\label{se:intro}

The Barrett-Crane model of four-dimensional Riemannian quantum 
gravity~\cite{BC} has been of significant interest
recently~\cite{Baez,JB,Oriti,PR}.
The model is discrete and well-defined, and
the partition function for the Perez-Rovelli version has been rigorously 
shown to converge~\cite{P} for a fixed triangulation of spacetime.
The Riemannian model serves as a step along the way to understanding
the less tractable but physically more realistic Lorentzian version~\cite{BC2}.
However, despite its simplicity, we are currently lacking explicit
numerical computations of the partition function and of expectation
values of observables in the Riemannian model.
These are necessary to test its large-scale behaviour and other physical 
properties.

It has been shown~\cite{BaCh} that the amplitudes in the Barrett-Crane
model are always non-negative, and therefore that the expectation
values of observables can be approximated using the Metropolis
algorithm.  This greatly reduces the number of samples that must
be taken, and thus the remaining obstacle is the time required to
compute each sample.  This paper presents a very efficient algorithm
for doing these computations.  
The algorithm is used in~\cite{BCE} and~\cite{BCHT} to understand
the asymptotic behaviour of the $10j$ symbols and the dependence of 
the partition function on a cutoff.

To explain further, we need to describe the Barrett-Crane model in
more detail.
It has been formulated by Baez~\cite{JB} as a discrete spin foam model, in which
faces in the dual 2-skeleton of a fixed triangulation of spacetime are labeled by spins.
The dual 2-skeleton consists of a dual vertex at the center of each 4-simplex of the
triangulation, five dual edges incident to each dual vertex (one for each tetrahedron in
the boundary of the 4-simplex), and ten dual faces incident to each dual vertex (one for
each triangle in the boundary of the 4-simplex).

Baez notes that the partition function for this model is the sum, over all labelings of
the dual faces by spins, of an expression that contains the product of a $10j$ symbol for 
each dual vertex.
A $10j$ symbol, described in detail in Section~\ref{se:10j}, is 
a \Spinf spin network. 
Roughly speaking, a spin network is a graph whose vertices are
labelled by tensors, and whose edges indicate how to contract these
tensors.
A spin network evaluates to a complex number in the
way explained in Section~\ref{se:elementary}.
In short, the $10j$ symbol 
is a function taking ten input spins and producing a complex number.
It is at the heart of the calculation of the partition function, and thus
an algorithm for calculating the $10j$ symbols efficiently is quite important.

In Section~\ref{se:10j} we recall the definition of the $10j$ symbol
using spin networks.
Then in Section~\ref{se:elementary} we briefly describe the elementary
algorithms for evaluating these spin networks, and give their running times
and memory use.
We conclude with Section~\ref{se:ours}, which presents our algorithms
and their time and space needs.

\section{The $10j$ symbol}\label{se:10j}

In the dual 2-skeleton of a triangulation of a 4-manifold, each dual vertex
belongs to five dual edges, and each pair of these dual edges borders a
dual face.
A $10j$ symbol is a \Spinf spin network with five vertices (corresponding to the
five dual edges) and ten edges, one connecting each pair of vertices (corresponding
to the ten dual faces), with the edges labeled by spins.
In the context of a \Spinf spin network,
a spin $j$ labeling an edge denotes the representation $j \otimes j$ of
$\Spinf \cong \SUt \times \SUt$,
where $j$ is the spin-$j$ representation of \SUt.
Such representations are called ``balanced.''
We use the convention that spins are non-negative half-integers.

Here is a picture of the $10j$ symbol, with the vertices numbered 0 through 4,
and the spins divided into two groups:  $j_{1,i}$ are the spins on the edges 
joining vertex $i$ to vertex $i+1$ (modulo 5), and $j_{2,i}$ are the spins
on the edges joining vertex $i$ to vertex $i+2$ (modulo 5).
\begin{equation}\label{E:10j}
\Center{\includegraphics{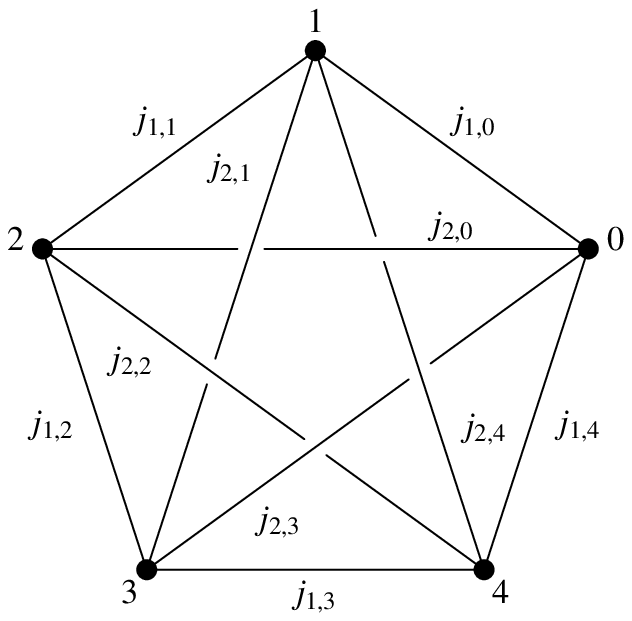}}
\end{equation}
The five vertices of the network are equal to Barrett-Crane intertwiners.
These are the unique intertwiners (up to a factor) between four balanced representations
of \Spinf with the property that their expansion as a sum of tensor products
of trivalent \SUt networks only
contains balanced representations on the internal edge, regardless of which pairs of
external edges are joined~\cite{MR}.
Barrett and Crane give the formula for these intertwiners in~\cite{BC}:
\[
\FourX{j_1}{j_2}{j_3}{j_4} \, := \, \sum_l \, \Delta_l \,
\DoubleY{j_1}{j_2}{j_3}{j_4}{l}
\quad
\DoubleY{j_1}{j_2}{j_3}{j_4}{l}
\]
Here the sum is over all admissible values of
$l$, i.e.\ those that satisfy the Clebsch-Gordan condition for both \SUt vertices.
So $l$ ranges from $\max(|j_1-j_2|,|j_3-j_4|)$ to $\min(j_1+j_2,j_3+j_4)$
in integer steps.  If the difference between these bounds is not an integer, 
the \Spinf vertex will be zero.
When $l$ satisfies these conditions, there is a unique intertwiner up to
normalization which can be used to label the trivalent \SUt vertices.
(These intertwiners are normalized so that the theta network in the
numerator of equation~(\ref{E:phi}) has value 1.)
$\Delta_l$ is the value of a loop in the spin-$l$ representation, which is just
$(-1)^{2l} (2l+1)$, the superdimension of the representation.

The uniqueness result of Reisenberger~\cite{MR} tells us that if
we replace the vertical edges in the above definition by horizontal
edges, the result differs at most by a constant factor.  In fact,
Barrett and Crane~\cite{BC} stated that the two definitions give exactly
the same \Spinf vertex, and Yetter~\cite{DY} has proved this.

Any closed spin network evaluates to a complex number, by contracting
the tensors at the vertices according to the pairings specified by the
edges. 
Thus the $10j$ symbol is a complex number.
(In fact, one can show that it is always a real number.)

To avoid confusion, we want to make it clear that we are working
with the ``classical'' (non $q$-deformed) evaluation of our
spin networks.  We will frequently reference the book~\cite{KL}
by Kauffman and Lins;  while it explicitly discusses the $q$-deformed
version, the formulas we use apply to the classical evaluation as well.

\section{Elementary algorithms}\label{se:elementary}

To set the context, we begin by explaining some elementary algorithms for
computing the $10j$ symbol.  These algorithms all share the feature that
they evaluate a spin network by choosing bases for the representations
labelling the edges, computing the components of the tensors representing
the intertwiners, and computing the contraction of the tensors in some
way.  They make no use of special features of these tensors, except for
the vanishing property mentioned below.

The first three methods each have two versions, one which works directly with
the \Spinf network~(\ref{E:10j}), and one that converts it into
a five-fold sum over \SUt networks, by expanding each 
Barrett-Crane intertwiner:
\begin{equation}\label{E:decagonal}
\sum_{l_0, \ldots, l_4} \,\, (\prod_{k=0}^{4} \, \Delta_{l_{k}}) \,\,
\Center{\includegraphics{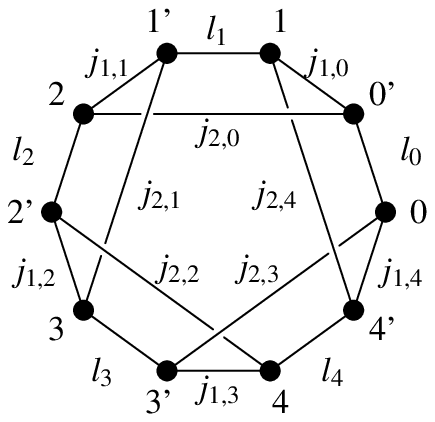} \, \includegraphics{10j_Fig3.eps}}
\end{equation}
Here, $l_i$ is the spin labeling the new edge introduced by the expansion of the
intertwiner at vertex $i$, and it ranges in integer steps from
$L_{i} := \max(|j_{1,i}-j_{2,i}|,|j_{1,i-1}-j_{2,i-2}|)$ to
$H_{i} := \min(j_{1,i}+j_{2,i},j_{1,i-1}+j_{2,i-2})$,
where the vertex numbers are all to be interpreted modulo 5. 
(If $H_i-L_i$ is not a non-negative integer, then the sum over $l_{i}$
is empty and the $10j$ symbol is zero.
In fact, if vertex $i$ is non-zero, then 
$|j_{1,i}-j_{2,i}|$, $|j_{1,i-1}-j_{2,i-2}|$,
$j_{1,i}+j_{2,i}$ and $j_{1,i-1}+j_{2,i-2}$ must all differ by integers.)
There are at most $\order(j)$ terms in each sum, where $j$ is the
average of the ten spins.
Since the two decagonal networks are the same, one only needs to evaluate
one of them and square the answer.

The first elementary method is one we call \textbf{direct contraction}.
One simply labels each edge in the spin network with a basis
vector from the representation labelling the edge, and multiplies
together the corresponding components of the tensors.
Then this is summed up over all labellings.
In fact, one can restrict to a smaller set of labellings:
the bases can be chosen so that
for each choice of two basis vectors on two of the
three edges meeting an \SUt vertex, there is at most one
choice of basis vector on the third edge giving a non-zero
tensor component.  The \Spinf vertices also have the property
that the bases can be chosen so that 
when three of the basis vectors adjacent to a vertex are specified, 
the last one is determined.

The second elementary method is \textbf{staged contraction}.
In the \Spinf version of this method, one starts with the tensor at vertex 0,
contracts with the tensor at vertex 1, and then vertex 4,
and then vertex 2, and finally vertex 3, again 
taking care to save space and time by using the
vanishing properties of the tensors.
Similarly, one can iteratively contract the tensors in the decagonal 
\SUt network.
At intermediate stages one is storing tensors with a large
number of components.

The third elementary method is \textbf{3cut}. 
Here one takes a ray from the center of~(\ref{E:10j}) and cuts
the three edges it crosses.
Then one takes the trace of the operator this defines
on the three-fold tensor product.
In more detail, one sums over basis vectors for the factors in this tensor
product, computing the effect of the network on these basis
vectors, and using the vanishing properties.
The memory required for this method (and the next) is dominated 
by the memory needed to store the tensors themselves.

The fourth and final elementary method is \textbf{2cut}.
This one only makes sense for the decagonal network, since
one proceeds by taking a ray from the center of the decagon
which crosses just two edges, cutting those two edges, and
taking the trace of the resulting operator, using the
vanishing properties.

Here is a table which gives an upper bound on the number of 
operations (additions and multiplications) that these algorithms
use, and the amount of memory they require, as a function
of a typical spin $j$.  The space requirements include the
space to store the Barrett-Crane tensors.

\bigskip
\begin{center}
\begin{tabular}{|c|c|c|c|c|c|}
\hline
       &       & direct contraction & staged contraction & 3cut        & 2cut    \\
\hline
\Spinf & Time  &  $j^{12}$          &  $j^{12}$          & $j^{12}$    & N/A     \\
       & Space &  $j^{6}$           &  $j^{10}$          & $j^{6}$     & N/A     \\
\hline
\SUt   & Time  &  $j^{11}$          &  $j^{9}$           & $j^{10}$    & $j^{9}$ \\
       & Space &  $j^{2}$           &  $j^{4}$           & $j^{2}$     & $j^{2}$ \\
\hline
Either & Time  &  $m p^6$           &  $m p^{2v+4}$      & $m p^{v+5}$ & $m p^4$ \\
       & Space &  $p^{v+2}$         &  $p^{v+4}$         & $p^{v+2}$   & $p^2$   \\
\hline
\end{tabular}
\end{center}
\bigskip

In the last two rows, we represent the entries in a uniform way
for either version by writing $v = 1$ for the \Spinf
version of each algorithm and $v = 0$ for the \SUt version.
Then we let $p = j^{v+1}$ and $m = j^{5-5v}$.
This shows how they are related.  For example, the \SUt methods
always get a factor of $j^{5}$ in time from the five loops coming
from expanding the Barrett-Crane vertices.
Also, the \Spinf methods get powers of $j^{2}$ 
(because $\dim j \otimes j = (2j+1)^2$), 
while the \SUt methods get powers of $j$
(because $\dim j = (2j+1)$).

The 2cut method has the best worst-case behaviour, in both space and time.

In the next section we present an algorithm which has running time
$\order(j^{5})$ and requires $\order(j^{2})$ space, and give variants
with running time $\order(j^{6})$ and $\order(j^{7})$ and which use
a constant amount of space.

The difference between $j^{5}$ and the running time, $j^{9}$\!,
for the best of the elementary methods is significant.
For example, with all spins equal to 20, our $j^{5}$ algorithm runs
in under six minutes on a 300 MHz microprocessor.  A back of the
envelope calculation suggests that this would take about
30 years with a $j^{9}$ algorithm.

\section{Our new algorithms}\label{se:ours}

In this section we describe a new method for computing $10j$ symbols.
The key feature of this method is that it does not proceed by
computing the tensor components for the intertwiners.  Instead
it uses recoupling to simplify the network to one that can be
evaluated directly.  Thus this method makes use of special
properties of the tensors that occur in the $10j$ symbols.

There are three versions of this method.  We explain one of these in
some detail, and briefly describe the variants
at the appropriate points. 
We use the notation of equation~(\ref{E:10j}), and consider the
expansion~(\ref{E:decagonal}) as a sum of squares of decagonal networks.

The decagonal networks can be deformed into the ``ladders'' shown below, where the
vertices at the bottom of each ladder are to be identified with those at the top of
the same ladder.  Any sign introduced by this deformation is produced twice and
so can be ignored.
\[
\sum_{l_0, \ldots, l_4} \,\, (\prod_{k=0}^{4} \,\Delta_{l_{k}})\,
\Center{\includegraphics{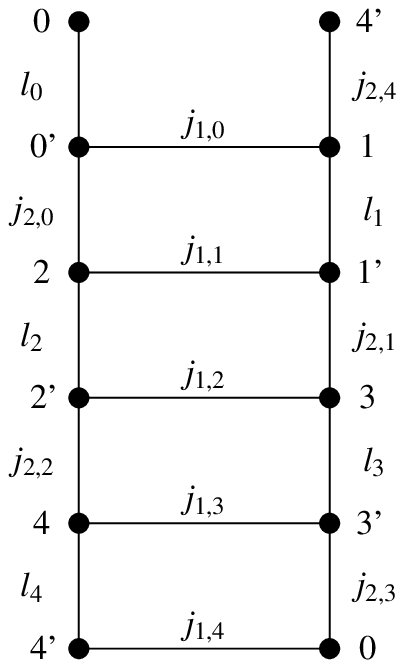}} \quad \Center{\includegraphics{10j_Fig4.eps}}
\]
To simplify these networks further, we recouple the sub-networks consisting of a
horizontal edge and half of each vertical edge incident to its endpoints,
rewriting them as sums of sub-networks with the same external edges, using
the following recoupling formula for $\SUt$ spin networks~\cite[Ch.~7]{KL}:
\[
\DoubleYhor{b}{c}{a}{d}{j} = \sum_{k} \,\,
\sixJ{a & b & k \cr
      c & d & j}
\DoubleY{b}{c}{a}{d}{k}
\]
The horizontal and vertical networks appearing above 
are two different ways
of writing intertwiners from one two-fold tensor product of irreducible representations
to another; both kinds of networks form bases for the space
of intertwiners, and the $6j$ symbols appearing in the formula
are defined to be the change-of-basis coefficients.

At first glance, it might look as if this recoupling would introduce sums over ten
new spins, labeling the ten new vertical edges.  However, the
total networks that result from the recoupling consist of chains of
sub-networks shaped like
\begin{equation}\label{E:phi1}
\xyphi{a}{b}{c}{d}
\end{equation}
By Schur's Lemma, such sub-networks will only be non-zero when the incoming and outgoing
edges have identical spins.  So the recoupled networks can
be written as a sum over just two new spins, $m_1$ and $m_2$.  
\[
\sum_{l_0, \ldots, l_4} \,\, (\prod_{k=0}^{4} \, \Delta_{l_{k}})\, \sum_{m_1,m_2} \,
\begin{matrix}
  \vbox to 200 pt{
    \vfill
    \hbox{\sixJ{j_{2,0} & l_0 & m_1 \cr
             j_{2,4} & l_1 & j_{1,0}}} 
    \vfill
    \hbox{\sixJ{l_2 & j_{2,0} & m_1 \cr
                l_1 & j_{2,1} & j_{1,1}}}
    \vfill
    \hbox{\sixJ{j_{2,2} & l_2 & m_1 \cr
                j_{2,1} & l_3 & j_{1,2}}}
    \vfill
    \hbox{\sixJ{l_4 & j_{2,2} & m_1 \cr
                l_3 & j_{2,3} & j_{1,3}}}
    \vfill
    \hbox{\sixJ{j_{2,4} & l_4 & m_1 \cr
                j_{2,3} & l_0 & j_{1,4}}}
    \vfill
  }
  \hbox{\includegraphics{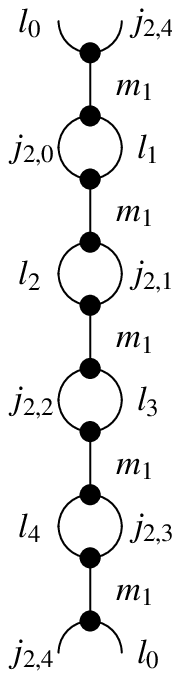}}
  \quad 
  \vbox to 200 pt{
    \vfill
    \hbox{\sixJ{j_{2,0} & l_0 & m_2 \cr
                j_{2,4} & l_1 & j_{1,0}}}
    \vfill
    \hbox{\sixJ{l_2 & j_{2,0} & m_2 \cr
                l_1 & j_{2,1} & j_{1,1}}}
    \vfill
    \hbox{\sixJ{j_{2,2} & l_2 & m_2 \cr
                j_{2,1} & l_3 & j_{1,2}}}
    \vfill
    \hbox{\sixJ{l_4 & j_{2,2} & m_2 \cr
                l_3 & j_{2,3} & j_{1,3}}}
    \vfill
    \hbox{\sixJ{j_{2,4} & l_4 & m_2 \cr
                j_{2,3} & l_0 & j_{1,4}}}
    \vfill
  }
  \hbox{\includegraphics{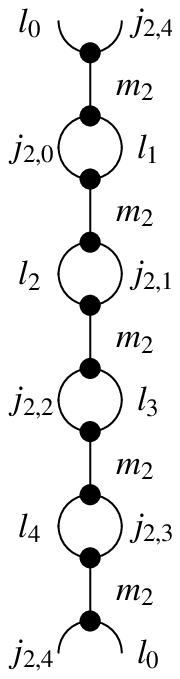}}
\end{matrix}
\]
This sum
is over all values such that every vertex in the diagram satisfies the
Clebsch-Gordan condition, so both $m_1$ and $m_2$ will independently 
range in integer steps from
$\max_{i}(|l_i-j_{2,i-1}|)$ to $\min_{i}(l_i+j_{2,i-1})$.
Here we use the fact that if the five Barrett-Crane vertices are non-zero,
then as $i$ varies, the ten quantities $|l_{i}-j_{2,i-1}|$ and $l_i+j_{2,i-1}$
all differ by integers. 
Indeed, $|l_{i}-j_{2,i-1}| \equiv l_{i}-j_{2,i-1} \equiv l_i+j_{2,i-1}$ modulo
integers, and by the paragraph after equation~(\ref{E:decagonal}),
\[
       l_i               + j_{2,i-1}
\equiv |j_{1,i}-j_{2,i}| + j_{2,i-1} 
\equiv j_{1,i}   + j_{2,i-1} + j_{2,i}
\equiv j_{1,i+1} + j_{2,i+1} + j_{2,i}
\equiv l_{i+1}               + j_{2,i}
\]
modulo integers, where in the third step we use that vertex $i+1$ is non-zero.

At this point there is a choice which determines which version of the
algorithm one obtains.
If the sum over $m_{1}$ and $m_{2}$ is left inside the sum over the $l_{i}$,
then it can be written as the square of a sum over a single $m$.
As described below, each of the terms in this sum can be computed
with $\order(j)$ operations
and a constant amount of memory, where $j$ is the average of the ten spins.
Thus, this method produces an algorithm that runs in $\order(j^{7})$ time
and takes a constant amount of space.

In general it turns out to be more efficient
to make the sum over the $m$'s outermost, in order to reinterpret
the sum over the $l_i$ as the trace of a matrix product.  
The range for $m_1$ and $m_2$ must encompass all potentially
admissible values, and the range for each $l_i$ can then be adjusted for the current
values of $m_1$ and $m_2$. The original range for $l_i$ was from $L_i$ to $H_i$
(see the paragraph after equation~(\ref{E:decagonal})),
so the $m$'s can never be greater than $\min_{i}(H_i+j_{2,i-1})$ without violating the
triangle inequality at one of the vertices.  The lower bound is given by
$\max_i(\min_{l_i}(|l_i-j_{2,i-1}|:L_i\leq l_i\leq H_i))$, where the minimum breaks down into
three cases:
if $j_{2,i-1}\geq H_i$, it is $j_{2,i-1}-H_i$;
if $j_{2,i-1}\leq L_i$, it is $L_i-j_{2,i-1}$; and
if $L_i<j_{2,i-1}<H_i$, it is either 0 or $\frac{1}{2}$, depending on whether
$2j_{2,i-1} \equiv 2L_i \pmod{2}$, or not.

Each of the $l_i$ is then restricted to take account of the current $m$ values,
ranging from
$\max(L_i,|m_1-j_{2,i-1}|,|m_2-j_{2,i-1}|)$ to
$\min(H_i, m_1+j_{2,i-1},  m_2+j_{2,i-1})$.

By Schur's Lemma, each sub-network of the form~(\ref{E:phi1}) is 
equal to a multiple of the identity,
so each of the recoupled ladders, with top and bottom edges joined, is simply a
multiple of a loop.  Kauffman and Lins~\cite{KL} give the following formula,
which can be checked by taking the trace of both sides:
%
%
\begin{equation}\label{E:phi}
\xyphi{a}{b}{c}{a}
=
\frac{
\begin{xy}
\xygraph{!{<0.8pc,0pc>:}
   [r] *{\bullet}
   -@-_{b} [ll] [rr]
   -@/_1.1pc/_{a} [ll] [rr]
   -@/^1pc/^{c} [ll] 
   *{\bullet}}
\end{xy}
}
{ \quad   
\begin{xy}
*{\xycircle<-.8pc,-.8pc>{}}, (5,-2) *{a} 
\end{xy}
} 
\quad
\begin{xy}
\xygraph{!{<2pc,0pc>:}
   [u]
   -@-^{a} [dd]}
\end{xy}
\end{equation}
The numerator of the fraction is written $\theta(a,b,c)$,
and the denominator is $\Delta_{a} = (-1)^{2a} (2a+1)$, 
the superdimension of the representation.
Kauffman and Lins also give a formula for the $6j$ symbol in terms of 
tetrahedral and $\theta$ networks:
\begin{equation}
\label{E:sixJ}
\sixJ{a & b & i \cr c & d & j} =
\frac{\Tet{ a & b & i \cr c & d & j} \Delta_i}
 {\theta(a,d,i)\,\theta(b,c,i)}
\end{equation}
Equations~(\ref{E:phi})
and~(\ref{E:sixJ}) allow us to write the following expression for the $10j$ symbol:
\begin{equation}
\label{E:tenJ}
\sum_{m_1,m_2}(2m_1+1)(2m_2+1)(-1)^{2(L_0+j_{2,4})-m_1-m_2}
\sum_{l_0, \ldots, l_4}\,\prod_{k=0}^{4}\,(M_k^{m_1,m_2})_{l_k}^{l_{k+1}}
\end{equation}
where
\begin{equation}
\label{E:matrix}
 (M_k^{m_1,m_2})_{l_k}^{l_{k+1}}=
 \frac{\Delta_{l_{k}} \,
 \Tet{ l_k & j_{2,k} & m_1 \cr l_{k+1} & j_{2,k-1} & j_{1,k}} \,
 \Tet{ l_k & j_{2,k} & m_2 \cr l_{k+1} & j_{2,k-1} & j_{1,k}}}
 {\theta(j_{2,k},l_{k+1},m_1)\,\theta(j_{2,k},l_{k+1},m_2)}
\end{equation}
The twists implicit in the identification of the top and bottom parts of each network
introduce signs of $(-1)^{l_0+j_{2,4}-m_1}$ and $(-1)^{l_0+j_{2,4}-m_2}$;
since $2l_0 \equiv 2L_0 \pmod{2}$, the product is $(-1)^{2(L_0+j_{2,4})-m_1-m_2}$.
The loop values for the spin-$m_1$ and spin-$m_2$ representations have signs of
$(-1)^{2m_1}$ and $(-1)^{2m_2}$, but since $2m_1 \equiv 2m_2 \pmod{2}$, the
product of these two signs is always unity.
We have made use of the symmetries of the tetrahedral networks to put the coefficients
in a uniform order for all terms.

The sum over the $l_i$ in Equation~(\ref{E:tenJ}) is the trace of the product of the
five matrices $M_k^{m_1,m_2}$.  For each pair of values of $m_1$ and $m_2$,
these matrices can be computed using closed formulas for the tetrahedral
and $\theta$ networks given by Kauffman and Lins in~\cite{KL}.  The formula for the
tetrahedral networks involves a sum with $\order(j)$ terms, so computing each matrix
requires $\order(j^3)$ operations\footnote{Each of the $\order(j)$ terms 
in the formula for the tetrahedral network 
contains factorials, which themselves require $\order(j)$ operations.  
However, with some care, the formula can be evaluated with a total of 
$\order(j)$ operations.  
In practice, we precalculate the factorials, using $\order(j)$ space.}.
The trace of the matrix product can also be found
in $\order(j^3)$ steps.  There are two factors of $j$ coming from
the sums over $m_1$ and $m_2$, yielding
an overall count of $\order(j^5)$ operations.
This method requires $\order(j^{2})$ space to store the matrices.

For some 10-tuples of spins, if all the spins are multiplied by $\lambda$, the time
required will scale at a lower power than $\lambda^5$.
Multiplying all the spins by a factor will increase the upper and lower bounds of all
the sums linearly, but in cases where the two bounds are equal, the sum will
consist of a single term, regardless of the scaling factor.
When many of the upper and lower bounds coincide, the first variant of the algorithm,
with worst case running time $\order(j^{7})$, in fact becomes faster than
the $\order(j^{5})$ version.
Thus one may wish to use the first variant for certain $10j$ symbols. 

For large spins, the memory usage can be a problem.  
For example, with spins of around 180, storing each matrix
$M_k^{m_1,m_2}$ requires about 1 gigabyte.
In this case, one can recalculate the matrix entries as needed,
resulting in $\order(j^{6})$ time and $\order(j^{0})$ space
($\order(j^{1})$ if factorials are cached).

The formulas in~\cite{KL} for the network evaluations are unnormalized.
To normalize all the \SUt intertwiners according to the convention that any $\theta$
network has a value of 1---which is the convention used in the formula for the
Barrett-Crane intertwiner---it is simpler to divide the matrix
elements by the appropriate $\theta$ networks than to take the existing $\theta$
networks in Equation~(\ref{E:matrix}) to be unity and normalize the tetrahedral
networks.  Including this normalization, the matrices become:
\begin{equation}
\label{E:norm}
 (N_k^{m_1,m_2})_{l_k}^{l_{k+1}}=
 \frac{(M_k^{m_1,m_2})_{l_k}^{l_{k+1}}}
 {\theta(j_{2,k-1},l_{k+1},j_{1,k})\,\theta(j_{2,k+1},l_{k+1},j_{1,k+1})
 }
\end{equation}

A subroutine written in C++ that implements this algorithm is available on the
web~\cite{DC}, along with some sample computations.

We have not dealt explicitly with the $q$-deformed case, where the representations of
\SUt are replaced with representations of $\SUt_q$, but it is straightforward
to adapt each stage of the development above, using the formulas in~\cite{KL} for the
$q$-deformed twist, loop, $\theta$ and tetrahedral networks.


\begin{thebibliography}{99}
\bibitem{Baez} J.C.\ Baez, Spin foam models, 
{\sl Class.\ Quantum Grav.\ }{\bf 15} (1998), 1827--1858. 

\bibitem{JB}
J.C. Baez,
An introduction to spin foam models of quantum gravity and BF theory,
in \emph{Geometry and Quantum Physics},
edited by Helmut Gausterer and Harald Grosse,
Springer, Berlin, 2000.
Preprint available as gr-qc/9905087.

\bibitem{BaCh}
J.C. Baez and J.D. Christensen,
Positivity of spin foam amplitudes,
to appear in \emph{Class.\ Quantum Grav.}
Preprint available as gr-qc/0110044.

\bibitem{BCE}
J.C. Baez, J.D. Christensen and G. Egan,
Asymptotics of $10j$ symbols.
In preparation.

\bibitem{BCHT} J.C. Baez, J.D. Christensen, T. Halford and D. Tsang, 
Spin foam models of Riemannian quantum gravity.
In preparation.

\bibitem{BC}
J. Barrett and L. Crane,
Relativistic spin networks and quantum gravity,
\emph{J. Math. Phys.} \textbf{39} (1998), 3296--3302.
Preprint available as gr-qc/9709028.

\bibitem{BC2} J.W.\ Barrett and L.\ Crane, A Lorentzian signature 
model for quantum general relativity, {\sl Class.\ Quantum 
Grav.\ }{\bf 17} (2000), 3101--3118. 

\bibitem{DC}
J.D. Christensen,
Spin foams page,
\texttt{http://jdc.math.uwo.ca/spin-foams/}.

\bibitem{KL}
L. Kauffman and S. Lins,
\emph{Temperley-Lieb recoupling theory and invariants of 3-manifolds},
Princeton University Press, Princeton, 1994.

\bibitem{Oriti} D.\ Oriti, Spacetime geometry from algebra: spin foam
models for non-perturbative quantum gravity, available as
gr-qc/0106091.

\bibitem{P} A.\ Perez, Finiteness of a spin foam model for
Euclidean quantum general relativity, {\sl Nucl.\ Phys.\ }{\bf B599}
(2001) 427--434.

\bibitem{PR} A.\ Perez and C.\ Rovelli, A spin foam model without
bubble divergences, {\sl Nucl.\ Phys.\ }{\bf B599} (2001) 255--282.

\bibitem{MR}
M.P. Reisenberger,
On relativistic spin network vertices,
\emph{J. Math. Phys.} \textbf{40} (1999), 2046--2054.
Preprint available as gr-qc/9809067.

\bibitem{DY}
D.N. Yetter,
Generalized Barrett-Crane vertices and invariants of embedded graphs,
\emph{J. Knot Theory Ramifications} \textbf{8} (1999), 815--829.
Preprint available as math.QA/9801131.

\end{thebibliography}
\end{document}